\documentclass[cameraready]{Interspeech}

\usepackage{overpic}
\usepackage{enumitem} 
\usepackage{overpic} 
\usepackage{color}
\usepackage[dvipsnames]{xcolor}
\usepackage{multirow,multicol}
\usepackage{xspace}

\usepackage{pifont}
\usepackage{amsfonts}
\usepackage{booktabs}
\usepackage{tabularx}

\usepackage{graphicx}
\usepackage{subfig}
\usepackage{pgfplots, mathtools}
\pgfplotsset{compat=1.18}

\usepackage{caption}
\definecolor{turquoise}{cmyk}{0.65,0,0.1,0.3}
\definecolor{purple}{rgb}{0.65,0,0.65}
\definecolor{dark_green}{rgb}{0, 0.5, 0}
\definecolor{orange}{rgb}{0.8, 0.6, 0.2}
\definecolor{red}{rgb}{0.8, 0.2, 0.2}
\definecolor{darkred}{rgb}{0.6, 0.1, 0.05}
\definecolor{blueish}{rgb}{0.0, 0.3, .6}
\definecolor{light_gray}{rgb}{0.7, 0.7, .7}
\definecolor{pink}{rgb}{1, 0, 1}
\definecolor{greyblue}{rgb}{0.25, 0.25, 1}

\newcommand{\PreserveBackslash}[1]{\let\temp=\\#1\let\\=\temp}
\newcolumntype{C}[1]{>{\PreserveBackslash\centering}p{#1}}
\newcolumntype{R}[1]{>{\PreserveBackslash\raggedleft}p{#1}}
\newcolumntype{L}[1]{>{\PreserveBackslash\raggedright}p{#1}}

\renewcommand{\paragraph}[1]{\vspace{1pt}\noindent\textbf{#1}\textbf{.}}
\newcommand{\scriptnote}[1]{\footnote{\scriptsize{#1}}}

\newcommand{\eg}{\textit{e.g.,~}}

\newcommand{\shortname}{MixProLAP\xspace}

\newcommand{\furl}[1]{\scriptnote{\url{#1}}}

\newcommand{\xcaption}[1]{\vspace{0pt}\caption{#1}\vspace{0pt}}

\title{MixProLAP: Mixture-Induced Uncertainty Modeling \\ for Probabilistic Language-Audio Pretraining}

\author[affiliation={1}]{Yu}{Nakagome}
\author[affiliation={2}]{Jaesong}{Lee}
\author[affiliation={2}, correspondingauthor]{Soo-Whan}{Chung}

\address{
    $^1$ LINE WORKS Corporation, Japan \\
    $^2$ NAVER Cloud Corporation, South Korea
}
\email{yu.nakagome1220@gmail.com, soowhan.chung@navercorp.com}


\usepackage{comment}

\begin{document}

\maketitle

\begin{abstract}

Acoustic environments often contain multiple overlapping sound events, and the same acoustic scene can be described using diverse textual expressions, making audio-text alignment inherently ambiguous. This paper proposes a probabilistic audio-language pretraining framework to model many-to-many correspondence ambiguity in audio-text alignment. Unlike conventional contrastive methods that learn deterministic point embeddings, our approach represents each modality as a distribution and learns uncertainty-aware cross-modal alignment. Rather than relying on masking-based uncertainty simulation, we mix audio-text pairs to create overlapping sounds that better reflect real acoustic mixtures and capture semantic inclusion relations among sound events. We further introduce a multi-level inclusion loss to enforce representations consistent with these relations. Experiments on audio-text retrieval benchmarks show that the proposed method outperforms deterministic baselines.

\end{abstract}

\section{Introduction}

Humans naturally describe complex acoustic environments using language. 
Learning to align audio signals with textual descriptions has therefore become an important problem in multi-modal representation learning.
Recent progress in cross-modal learning has been driven by contrastive pretraining approaches that learn joint embeddings from paired data~\cite{nagrani2018seeing,chung2019perfect,radford2021learning,elizalde2023clap,tschannen2025siglip2}.
Contrastive Language-Image Pretraining (CLIP)~\cite{radford2021learning} demonstrated the effectiveness of this paradigm by learning shared representations from large-scale image-caption datasets.
Following this success, Contrastive Language-Audio Pretraining (CLAP)~\cite{elizalde2023clap,audioclip2022, 10095969,10448504} extended the framework to audio-language learning, enabling models to associate acoustic scenes with natural language descriptions.

Despite their success, these approaches represent each modality as a deterministic point embedding.
This representation assumes a one-to-one alignment between audio and text, overlooking the inherently many-to-many correspondence of audio-language alignment.
A single acoustic event may be described in multiple ways, while a general textual description can correspond to a wide variety of acoustic realizations.
By encoding each pair as an independent point in the embedding space, deterministic models fail to capture uncertainty and hierarchical semantic relationships.
For example, the description ``\textit{heavy rain}" semantically entails ``\textit{rain}", yet point embeddings cannot naturally represent such inclusion structures.

To address ambiguity in cross-modal representation learning, recent works have explored probabilistic embeddings~\cite{Upadhyay_2023_ICCV, chun2025prolip,chun2021pcme,chun2024improved,chun2025longprolip, chun2025multiplicity}.
ProLIP~\cite{chun2025prolip} introduced a probabilistic vision-language framework that represents each modality as a Gaussian distribution rather than a single point.
Its learning objective, Probabilistic Pairwise Contrastive Learning (PPCL) with Closed-form Sampled Distance (CSD)~\cite{chun2024improved}, measures distances between distributions to explicitly model uncertainty.
In addition, ProLIP introduced inclusion-based objectives that capture asymmetric semantic relationships between modalities.

However, directly applying ProLIP's masking strategy to audio is problematic due to the temporal and compositional nature of sound.
ProLAP~\cite{manabe2025prolap} extended ProLIP to the audio domain using spectrogram masking for intra-modal uncertainty modeling.
While the inclusion objective assumes that the original input is a semantic subset of the masked input, this assumption often fails for audio.
For transient events (\eg gunshots or dog barks), masking removes the defining acoustic cue entirely, violating the subset relationship.
Conversely, for ambient sounds (\eg rain or music), masking preserves global semantics, failing to create meaningful hierarchical variations.
To address this limitation, we model uncertainty through additive audio composition rather than information removal.

In this paper, we propose \shortname, a probabilistic contrastive learning framework tailored for the audio domain.
We retain the core components of ProLIP, including PPCL and the inclusion objectives, while modifying the inclusion mechanism to model the temporal and compositional nature of audio.
We introduce an audio mixing process that superimposes two sounds, constructing a composite sample containing both acoustic events.
This provides a structured approximation of semantic inclusion and enables consistent application of the inclusion loss in the audio domain.
For text inputs, we generate composite captions by concatenating two descriptions with simple conjunctions such as “and” or “while” to reflect co-occurring events.
We also introduce a multi-level inclusion loss conditioned on the mixing ratio, enforcing graded semantic inclusion to promote the structured learning of uncertainty.
Experiments on AudioCaps and ClothoV2 demonstrate that \shortname achieves improved zero-shot retrieval performance while providing uncertainty estimates.

\section{Probabilistic Representation Learning}

Most contrastive multi-modal learning frameworks represent each modality using deterministic point embeddings, which do not explicitly model uncertainty in cross-modal relationships. 
To address this limitation, probabilistic representations have been proposed.
PCME~\cite{chun2021pcme} and PCME++~\cite{chun2024improved} model each input as a Gaussian distribution rather than a single point. 
Let $x$ denote an input sample from a modality (\eg an audio signal or a text description).
An encoder $f(\cdot)$ maps the input to the parameters of a Gaussian distribution by predicting a mean vector $\mu(x)$ and a variance vector $\sigma(x)$, forming the embedding
$p(x) = \mathcal{N}(\mu(x), \text{diag}(\sigma(x)))$.
Following ProLIP~\cite{chun2025prolip}, $\sigma(x)$ denotes the diagonal variance vector rather than the standard deviation.
Similarity between two embeddings $p_i$ and $p_j$ can then be computed using the Closed-form Sampled Distance (CSD)~\cite{chun2024improved}:
\begin{equation}\label{eq:csd}
d_{CSD}(p_i, p_j) = ||\mu_i - \mu_j||^2_2 + ||\sigma_i + \sigma_j||_1.
\end{equation}
Building on this idea, ProLIP~\cite{chun2025prolip} replaces conventional contrastive learning with Probabilistic Pairwise Contrastive Learning (PPCL), which computes similarity between Gaussian embeddings. 
The PPCL loss is formulated as
\begin{equation}\label{eq:ppcl}
    \mathcal{L}_{\text{PPCL}}(p_i,p_j) = \log\Big(1 + e^{y_{i,j} \left( -a \left( \mu_i^\top \mu_j - \frac{1}{2} ||\sigma_i +\sigma_j||_1 \right) + b \right)}\Big)
\end{equation}
where $y_{i,j}$ indicates whether the pair is positive ($1$) or negative ($-1$), and $a$ and $b$ are learnable scaling and bias parameters.
By incorporating uncertainty into the similarity computation, PPCL enables more flexible modeling of many-to-many relationships between modalities.

In addition to probabilistic contrastive learning, ProLIP introduces an inclusion objective to capture asymmetric semantic relationships between modalities.
Since textual descriptions are typically more abstract than visual inputs, the model assumes that image embeddings should be contained within the corresponding text embeddings.
The inclusion loss $\mathcal{L}_{\text{Inc.}}$ encourages the distribution of one modality to be included in another:
\begin{equation}\label{eq:inclusion}
    \mathcal{L}_{\text{Inc.}}({Z}_1\subset{Z}_2) = \log\Big(1 + e^{-c\mathcal{H}({Z}_1\subset{Z}_2)}\Big),
\end{equation}
where $Z_1$ and $Z_2$ denote two Gaussian variables with density functions $p_1$ and $p_2$, and $\mathcal{H}(Z_1 \subset Z_2)$ is a hypothesis score defined as
\begin{equation}\label{eq:hypothesis}
\begin{split}
\mathcal{H}({Z}_1\subset{Z}_2)= & \log\int_{-\infty}^{\infty}p_1^2(x)p_2(x)dx\\
&-\log\int_{-\infty}^{\infty}p_2^2(x)p_1(x)dx.
\end{split}
\end{equation}
ProLIP further introduces a masking-based inclusion mechanism to model intra-modal uncertainty.
Random masking is applied to image patches and text tokens to generate semantically reduced inputs.
The model then enforces that the representation of the original input is contained within that of the masked input: $\mathcal{L}_{\text{mask}}=\mathcal{L}_{\text{Inc.}}\big(Z(x)\subset Z(\tilde{x})\big)$,
where $\tilde{x}$ denotes the masked version of $x$.

Following this paradigm, ProLAP~\cite{manabe2025prolap} extends probabilistic representation learning to the audio-language domain by adopting PPCL and the inclusion objectives.
ProLAP introduces spectrogram patch masking for audio and proposes additional objectives such as Hierarchical Inclusion Loss and Mask Repulsive Loss to improve probabilistic alignment.
However, masking-based strategies can be problematic for audio due to its temporal and compositional nature.
Masking may remove short but semantically critical acoustic events, breaking the assumed subset relationship between original and masked samples.
In other cases, masking affects only local regions while leaving the overall acoustic semantics unchanged, failing to produce meaningful hierarchical variations.

\begin{figure}[t]
    \centering\footnotesize
        \includegraphics[width=\columnwidth]{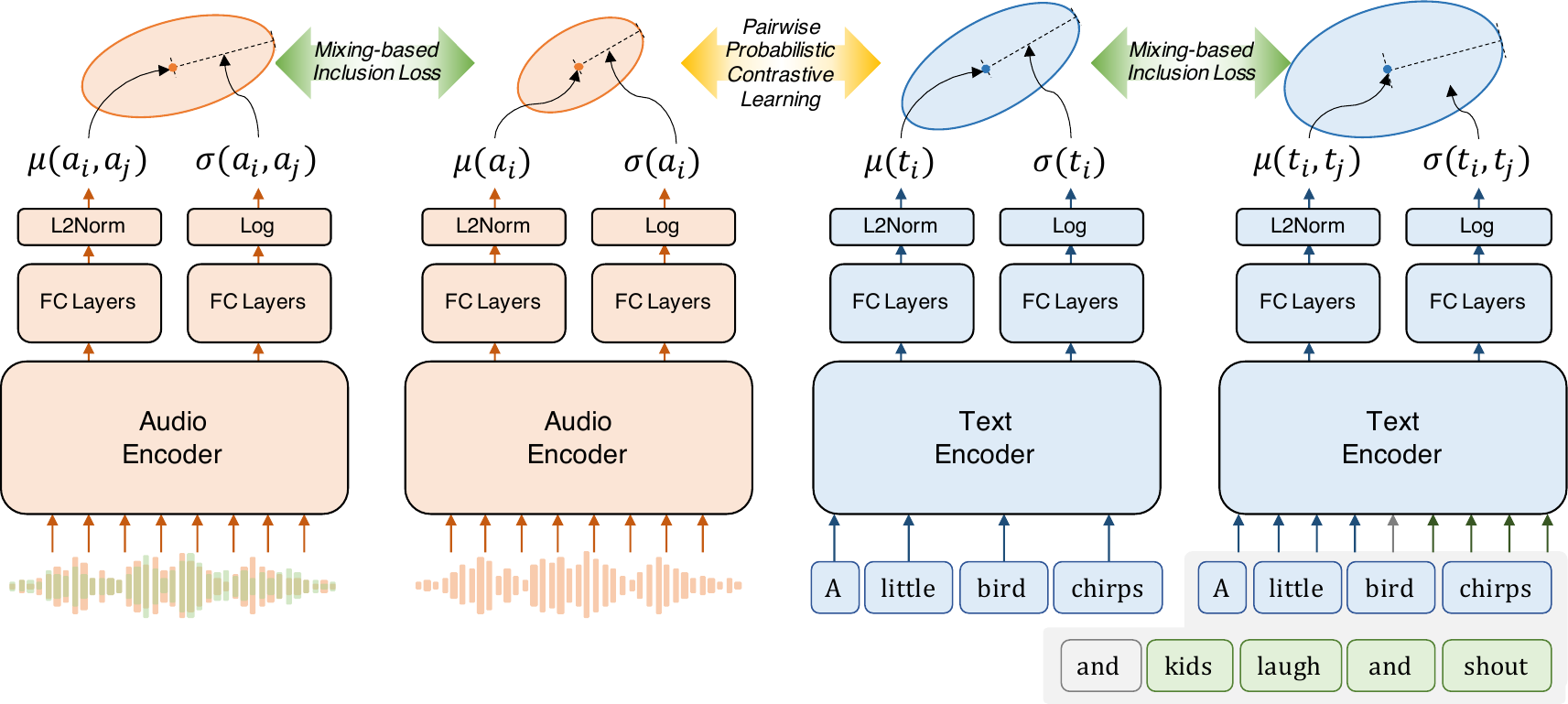}
\vspace{-6pt}
\xcaption{Overview of \shortname architecture. Audio and text encoders output probabilistic embeddings through mean and variance projection heads. PPCL aligns the distributions across modalities.}
\label{fig:block}
\vspace{-10pt}
\end{figure}

\section{Proposed Method: \shortname}

In this section, we introduce \shortname, a probabilistic framework that models uncertainty through audio mixing.
Our key contribution is replacing masking-based uncertainty with additive uncertainty through waveform mixing, enabling more appropriate hierarchical structure learning for audio signals.
An overview of the architecture is shown in Figure~\ref{fig:block}.

\subsection{Probabilistic Audio-Text Alignment}
We build \shortname on top of the CLAP architecture.
The audio encoder is HTS-AT~\cite{chen2022hts, liu2021swin} and the text encoder is GPT-2~\cite{radford2019language}.
To obtain probabilistic embeddings, we use two projection heads for each encoder to predict the mean and variance of a diagonal Gaussian distribution.
The mean embeddings are $L_2$-normalized to stabilize the latent space, while variances are predicted in log scale to ensure stability.

Following ProLIP, we align audio and text distributions using PPCL.
We also adopt an inter-modal inclusion objective that models semantic asymmetry between modalities by encouraging the audio distribution to be contained within the corresponding text distribution.
While these components provide the probabilistic foundation, the key challenge is how to induce meaningful uncertainty for audio representations.
We address this by introducing mixture-based uncertainty modeling.

\subsection{Mixture-induced Uncertainty Modeling}
Acoustic environments frequently contain multiple simultaneous sound events.
Unlike conventional mixing augmentation that increases data diversity, our mixing strategy explicitly constructs semantic superset relations between source signals and their mixtures, serving as a core component for inclusion-based probabilistic alignment learning.
To reflect this compositional nature, we construct mixture samples by combining two audio signals.
Given two audio waveforms $a_i$ and $a_j$, we generate a mixed signal
\begin{equation}
    a_{i,j} = \alpha a_i + (1 - \alpha) a_j,
\end{equation}
where the mixing coefficient is sampled from $\alpha \sim \mathcal{U}(0.5,1.0)$.
The pair $(a_i,a_j)$ is sampled from different audio-text pairs within the same minibatch.
The resulting mixture contains acoustic cues from both sources and can therefore be regarded as an approximate semantic superset of its components.

To maintain consistency between modalities, we construct corresponding textual mixtures.
Given captions $t_i$ and $t_j$, we generate a combined caption $t_{i,j} = \text{concat}(t_i,t_j)$, where captions are connected using simple conjunctions such as ``\textit{A, and B}'' or ``\textit{A while B}''.
This composition preserves the semantics of both captions while reflecting the increased informational content of the mixed audio signal.

\begin{figure}[t]
    \centering\footnotesize
    \begin{minipage}[t]{0.48\columnwidth}
        \includegraphics[width=\columnwidth]{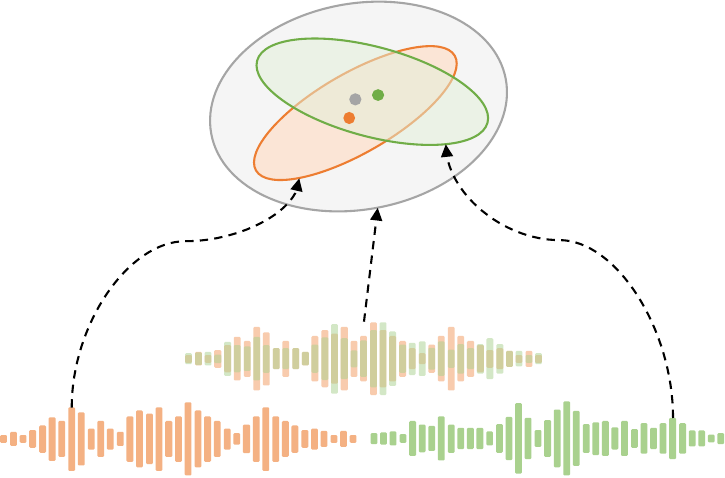}
        \centerline{(a) Audio Inclusion Loss}
    \end{minipage}\hfill
    \begin{minipage}[t]{0.48\columnwidth}
        \includegraphics[width=\columnwidth]{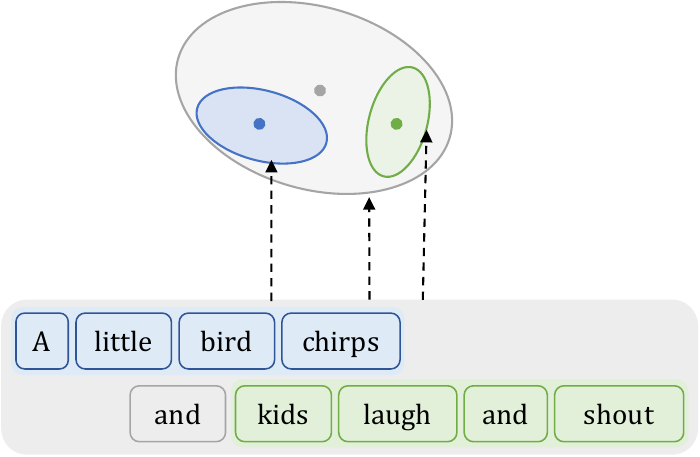}
        \centerline{(b) Text Inclusion Loss}
    \end{minipage}
\xcaption{Illustration of intra-modal inclusion loss. (a) Audio mixing creates a semantic superset where each source distribution is included within the mixture. (b) Text concatenation constructs a textual superset encompassing both captions.}
\label{fig:inclusion}
\vspace{-10pt}
\end{figure}

We enforce the inclusion relationship between sources and their mixtures as illustrated in Figure~\ref{fig:inclusion}.
For audio mixtures, we encourage the embeddings of the individual sources to be included within the mixture embedding:
\begin{equation}
\mathcal{L}_{\text{mix}}^{a}
=
\mathcal{L}_{\text{Inc.}}(Z(a_i)\subset Z(a_{i,j}))
+
\mathcal{L}_{\text{Inc.}}(Z(a_j)\subset Z(a_{i,j})).
\end{equation}
Similarly, for textual mixtures we enforce the inclusion as,
\begin{equation}
\mathcal{L}_{\text{mix}}^{t}
=
\mathcal{L}_{\text{Inc.}}(Z(t_i)\subset Z(t_{i,j}))
+
\mathcal{L}_{\text{Inc.}}(Z(t_j)\subset Z(t_{i,j})).
\end{equation}
These objectives encourage mixture representations to occupy broader regions in the embedding space while preserving the semantic relationships between individual and composite events.

\subsection{Multi-level Inclusion Loss}
Audio mixtures induce varying levels of semantic ambiguity depending on the mixing ratio.
To capture this property, we introduce a multi-level inclusion~(MLI) loss that enforces hierarchical relationships between mixtures with different compositions.
We define $L$ mixing levels with coefficients $\alpha_\ell$.
Each level generates a mixture as,
\begin{equation}\label{eq:hier_mix}
  a_{i,j}^{(\ell)} = \alpha_\ell a_i + (1-\alpha_\ell) a_j, \quad \alpha_0 > \alpha_1 > \cdots > \alpha_{L-1}.
\end{equation}
Higher levels correspond to mixtures that contain more balanced contributions from both sources and therefore greater semantic uncertainty.
The multi-level inclusion loss enforces hierarchical relationships between mixtures of increasing uncertainty:
\begin{equation}\label{eq:mliloss}
\mathcal{L}_{\text{MLI}}
=
\sum_{\ell=0}^{L-2}
\mathcal{L}_{\text{Inc.}}
\Big(
Z(a_{i,j}^{(\ell)}) \subset Z(a_{i,j}^{(\ell+1)})
\Big).
\end{equation}
This objective encourages a smooth gradient of uncertainty aligned with the mixing degree, promoting consistent hierarchical relationships across mixture levels.

\subsection{Training Objective}
The overall training objective combines the aforementioned loss terms to ensure robust probabilistic representation learning:
\begin{equation}
\label{eq:total_loss_}
\begin{aligned}
\mathcal{L}_{\text{total}} &=
\mathcal{L}_{\text{PPCL}}(Z(a_i),Z(t_i)) +
\lambda_{\text{inter}}\mathcal{L}_{\text{Inc.}}(Z(a_i)\subset Z(t_i))\\
&+
\lambda_a \mathcal{L}^{a}_{\text{mix}}(a_i,a_j)
+
\lambda_t \mathcal{L}^{t}_{\text{mix}}(t_i,t_j) \\
&+
\lambda_{\text{MLI}}\mathcal{L}_{\text{MLI}}(a_i,a_j)
+
\beta \mathcal{L}_{\text{VIB}}(Z(a_i),Z(t_i)),
\end{aligned}
\end{equation}
where $(a_i,t_i)$ is a paired audio-text sample, while $a_j$ and $t_j$ $(i \neq j)$ are negative samples randomly selected from the same batch.
$\mathcal{L}_{\text{VIB}}$ is the variational information bottleneck (VIB) loss that prevents variance collapse of probabilistic distribution~\cite{chun2025prolip,alemi2016vib}.

\section{Experiments}

\begin{table*}[th!]
    \centering
    \footnotesize
    \caption{
        Zero-shot audio-text retrieval results on AudioCaps and ClothoV2 test sets.
        AC: AudioCaps, CL: ClothoV2.
        The CLAP baseline is finetuned with InfoNCE loss using the same pretrained weights and training data as \shortname.
    }
    \vspace{-3pt}
    \label{tab:main}
    \setlength{\tabcolsep}{4pt}
    \begin{tabular}{llccccccccccccc}
        \toprule
        & & \multicolumn{6}{c}{AudioCaps} & \multicolumn{6}{c}{ClothoV2} \\
        \cmidrule(lr){3-8} \cmidrule(lr){9-14}
        & & \multicolumn{3}{c}{A$\rightarrow$T} & \multicolumn{3}{c}{T$\rightarrow$A} & \multicolumn{3}{c}{A$\rightarrow$T} & \multicolumn{3}{c}{T$\rightarrow$A} \\
        \cmidrule(lr){3-5} \cmidrule(lr){6-8} \cmidrule(lr){9-11} \cmidrule(lr){12-14}
        Method & Train & R@1 & R@10 & mAP@10 & R@1 & R@10 & mAP@10 & R@1 & R@10 & mAP@10 & R@1 & R@10 & mAP@10 \\
        \midrule
        CLAP~\cite{elizalde2023clap}             & AC & 24.23 & 63.71  & 18.89 & \textbf{26.85} & \textbf{71.44} & \textbf{39.93} & 11.67 & 36.36 & 8.36 & \textbf{14.56} & \textbf{45.55} & \textbf{23.18} \\
        \shortname        & AC & \textbf{26.85} & \textbf{68.37} & \textbf{20.24} & 25.53 & 70.63 & 38.76 & \textbf{14.26} & \textbf{46.89} & \textbf{9.96} & 12.33 & 41.47 & 20.06 \\
        \midrule
        CLAP~\cite{elizalde2023clap}             & CL & 19.68 & 55.52 & 13.40 & 20.20 & 61.07 & 31.65 & 13.40 & 41.72 & 9.65 & \textbf{16.56} & 48.86 & \textbf{25.11} \\
        \shortname         & CL & \textbf{19.80} & \textbf{60.86} & \textbf{14.99} & \textbf{21.05} & \textbf{63.82} & \textbf{33.16} & \textbf{15.60} & \textbf{46.51} & \textbf{11.19} & 15.62 & \textbf{50.01} & 25.08 \\
        \bottomrule
    \end{tabular}
\end{table*}

\begin{table}[t]
    \centering
    \footnotesize
    \caption{
        Ablation study on AudioCaps test set.
        \checkmark: enabled.
        Inc.: inter-modal inclusion loss.
        Mix.: mixing-based intra-modal inclusion loss.
        MLI: multi-level inclusion loss.
    }
    \vspace{-6pt}
    \label{tab:ablation}
    \setlength{\tabcolsep}{3pt}
    \begin{tabular}{cccccccc}
        \toprule
        & & & & \multicolumn{2}{c}{A$\rightarrow$T} & \multicolumn{2}{c}{T$\rightarrow$A} \\
        \cmidrule(lr){5-6} \cmidrule(lr){7-8}
        PPCL & Inc. & Mix. & MLI & R@1 & mAP@10 & R@1 & mAP@10 \\
        \midrule
        \checkmark  & \checkmark  &             &             & 22.98 & 18.55 & 26.05 & 39.26 \\
        \checkmark  & \checkmark  & \checkmark  &             & 24.69 & 19.34 & \textbf{26.67} & \textbf{39.91} \\
        \checkmark  & \checkmark  & \checkmark  & \checkmark  & \textbf{26.85} & \textbf{20.24} & 25.53 & 38.76 \\
        \bottomrule
    \end{tabular}
\end{table}

\begin{table}[t]
    \centering
    \footnotesize
    \caption{
        Comparison of intra-modal uncertainty strategies on AudioCaps test set.
    }
    \vspace{-6pt}
    \label{tab:mask_vs_mix}
    \setlength{\tabcolsep}{3pt}
    \begin{tabular}{llcccc}
        \toprule
        & & \multicolumn{2}{c}{A$\rightarrow$T} & \multicolumn{2}{c}{T$\rightarrow$A} \\
        \cmidrule(lr){3-4} \cmidrule(lr){5-6}
        Audio Aug. & Text Aug. & R@1 & mAP@10 & R@1 & mAP@10 \\
        \midrule
        Spec. mask    & Token mask     & 22.64 & 17.31 & 21.68 & 35.34 \\
        Mixing (Ours)     & Token mask & 23.89 & 17.11 & 19.80 & 33.12 \\
        Mixing (Ours)     & Concat.\ (Ours)   & \textbf{26.85} & \textbf{20.24} & \textbf{25.53} & \textbf{38.76} \\
        \bottomrule
    \end{tabular}
\vspace{-10pt}
\end{table}

\subsection{Experimental Settings}

\paragraph{Dataset and metrics}
We used AudioCaps~\cite{kim2019audiocaps} and ClothoV2~\cite{drossos2020clotho} datasets for both training and evaluation.
AudioCaps consists of approximately 51k audio clips from AudioSet~\cite{audioset} with human written captions, while ClothoV2 contains nearly 6k audio samples with five captions each.
All audio samples were processed as 10-second segments for training.
For the ClothoV2 test set, which contains longer audio clips, we split each audio into 10-second chunks and averaged their embeddings to obtain the final representation.
Following standard practice in audio-language retrieval~\cite{Primus2024}, we report Recall@1, Recall@10, and mAP@10 for both audio-to-text (A$\rightarrow$T) and text-to-audio (T$\rightarrow$A) retrieval.
Note that all five captions included in the testset were used as retrieval targets.
For probabilistic models, similarity between audio and text embeddings is computed using the CSD (Eq.~\eqref{eq:csd}) instead of cosine similarity.

\paragraph{Baselines}
We evaluated \shortname against a finetuned CLAP baseline model to ensure a fair comparison.
The baseline was initialized with the pretrained CLAP weights~\footnote{https://github.com/microsoft/CLAP} and finetuned on the same dataset using the standard InfoNCE loss~\cite{oord2018representation}.

\paragraph{\shortname Structure}
The audio encoder and text encoder were initialized with the same pretrained CLAP weights as the baseline model.
On top of each encoder, we attached two independent projection heads to predict the mean and variance of the Gaussian embedding.
Each projection head consists of two fully connected layers.
The mean head outputs are $L_2$-normalized to stabilize the latent space, while the variance head predicts log-variance to ensure numerical stability.
For the MLI loss (Eq.~\eqref{eq:mliloss}), we use $L=3$ mixture levels for audio.

\paragraph{Training Details}
Following ProLIP~\cite{chun2025prolip}, we set the initial bias of the variance head $b$ to $-10$.
This allows the model to start with very low uncertainty and gradually learn the uncertainty representation.
We optimized the model using AdamW~\cite{LoshchilovH19} with $\beta_1=0.9$, $\beta_2=0.999$, and a weight decay of $0.01$.
The learning rate was $1\times10^{-5}$ with cosine annealing scheduler and 5 epochs of linear warmup.
The model was trained for 30 epochs with an effective batch size of 2,048.
For the mixing strategy, the coefficient $\alpha_\ell$ is sampled uniformly from $(0.5, 1.0)$, with $50\%$ of each batch used for mixed samples.
Finally, we set the loss weights (Eq.~\eqref{eq:total_loss_}) as $\lambda_{\text{inter}}=5\times10^{-7}$, $\lambda_{a}=5\times10^{-3}$, $\lambda_{t}=5\times10^{-3}$, and $\beta=1\times10^{-5}$.

\subsection{Experimental Results}

Table~\ref{tab:main} compares \shortname against the deterministic CLAP baseline on zero-shot audio-text retrieval benchmarks.
We evaluate models trained on AudioCaps and ClothoV2 separately to assess both in-domain performance and out-of-domain generalization.
When trained on AudioCaps, \shortname consistently outperforms CLAP in A$\rightarrow$T retrieval across both evaluation datasets.
These results demonstrate that the probabilistic embeddings and mixing-based uncertainty strategy enable \shortname to capture the inherent ambiguity of audio-caption pairs more effectively.
However, for T$\rightarrow$A retrieval, CLAP shows slightly better performance on AudioCaps, while \shortname achieves competitive results on ClothoV2.
This asymmetric performance suggests that modeling uncertainty through simple text concatenation is less effective than audio mixing, which directly manipulates the acoustic signal structure.
When trained on ClothoV2, the advantage of \shortname becomes more pronounced.
\shortname achieves consistent improvements across both retrieval directions on in-domain evaluation.
More importantly, on out-of-domain AudioCaps evaluation, \shortname substantially outperforms CLAP, particularly in T$\rightarrow$A retrieval.
These results suggest that probabilistic modeling improves robustness to variations in caption expressions across datasets.

\begin{figure}[t]
\centering
\begin{minipage}[t]{\dimexpr(\linewidth-0.8mm)/2\relax}
  \centering
  \hspace*{-4mm}%
  \begin{tikzpicture}
    \begin{axis}[      
      width=1.2\linewidth,
      height=0.9\linewidth,
      xlabel={Audio Duration (seconds)},
      ylabel={Variance},
      xlabel style={font=\scriptsize},
      ylabel style={font=\scriptsize, yshift=-5pt, xshift=2pt},
      tick label style={font=\scriptsize},
      yticklabel style={xshift=1pt},
      xtick={2.5, 7.5, 12.5, 17.5, 22.5},
      y tick label style={/pgf/number format/fixed, /pgf/number format/precision=2},
      xmin=0, xmax=25,
      ymin=0, ymax=0.40,
      grid=major,
      grid style={gray!10},
    ]
    \addplot+[
      cyan,
      mark=*,
      mark options={fill=cyan},
      mark size=1pt,
      error bars/.cd,
      y dir=both,
      y explicit
    ]
    coordinates {
      (2.5,  0.0629) +- (0, 0.0000)
      (7.5,  0.2405) +- (0, 0.1290)
      (12.5, 0.1605) +- (0, 0.0872)
      (17.5, 0.2296) +- (0, 0.0884)
      (22.5, 0.2458) +- (0, 0.0903)
    };
    \end{axis}
  \end{tikzpicture}
  \centerline{\footnotesize (a) Audio Uncertainty}
\end{minipage}\hspace{0.8mm}%
\begin{minipage}[t]{\dimexpr(\linewidth-0.8mm)/2\relax}
  \centering
  \hspace*{-0.8mm}%
  \begin{tikzpicture}
    \begin{axis}[
      width=1.2\linewidth,
      height=0.9\linewidth,
      xlabel={Text Length (words)},
      ylabel={Variance},
      xlabel style={font=\scriptsize},
      ylabel style={font=\scriptsize, yshift=-5pt, xshift=2pt},
      tick label style={font=\scriptsize},
      yticklabel style={xshift=1pt},
      ytick={0.02, 0.025, 0.03}, 
      xtick={3, 8, 13, 18, 25},
      y tick label style={/pgf/number format/fixed, /pgf/number format/precision=2},
      xmin=0, xmax=28,
      ymin=0.015, ymax=0.035,
      grid=major,
      grid style={gray!10},
    ]
    \addplot+[
      orange,
      mark=*,
      mark options={fill=orange},
      mark size=1pt,
      error bars/.cd,
      y dir=both,
      y explicit
    ]
    coordinates {
      (3,  0.0284) +- (0, 0.0053)
      (8,  0.0235) +- (0, 0.0045)
      (13, 0.0218) +- (0, 0.0035)
      (18, 0.0209) +- (0, 0.0018)
      (25, 0.0185) +- (0, 0.0011)
    };
    \end{axis}
  \end{tikzpicture}
  \centerline{\footnotesize (b) Text Uncertainty}
\end{minipage}
\caption{
Relationship between input length and estimated uncertainty on AudioCaps test set.}
\label{fig:uncertainty}
\vspace{-15pt}
\end{figure}
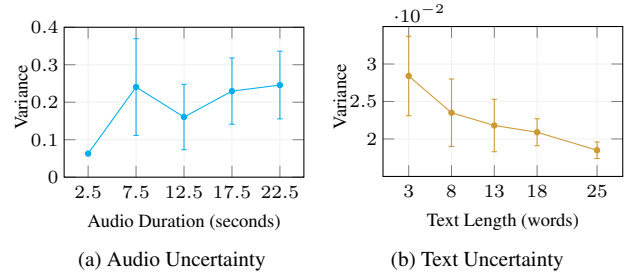

\subsection{Uncertainty Analysis}
To verify whether the variance embeddings effectively capture semantic density, we analyze the estimated uncertainty relative to input lengths in Figure~\ref{fig:uncertainty}. 
As expected, textual uncertainty decreases as caption length increases, indicating that richer descriptions provide clearer semantic cues and reduce ambiguity.
However, audio uncertainty remains relatively invariant to signal duration. 
Our analysis reveals that very short audio segments often contain distinct, well-defined acoustic events, resulting in low uncertainty.
In contrast, increasing the duration of a single audio clip does not result in additional semantic variety or diverse acoustic events; instead, it often results in temporal redundancy.

\subsection{Ablation Study}
Table~\ref{tab:ablation} reports an ablation study on AudioCaps to analyze the contribution of each loss term.
Starting from a baseline with PPCL and inter-modal inclusion loss (Inc.), we progressively add the mixing-based intra-modal inclusion loss (Mix.) and multi-level inclusion loss (MLI).
Adding Mix. improves performance in both retrieval directions.
Further adding MLI yields additional improvements in A$\rightarrow$T retrieval, while T$\rightarrow$A retrieval shows slight performance degradation.
Overall, the results indicate that the mixing-based inclusion loss primarily contributes to the observed improvements.

\subsection{Masking vs.\ Mixing}

Table~\ref{tab:mask_vs_mix} compares masking-based and mixing-based intra-modal uncertainty strategies on AudioCaps.
For the masking-based baseline, we follow the ProLAP~\cite{manabe2025prolap} setting.
Specifically, audio masking is implemented using SpecAug-style spectrogram masking~\cite{Park19_interspeech}, while text masking removes caption tokens.
We additionally adopt the hierarchical inclusion loss and mask repulsive loss proposed in ProLAP to ensure a fair comparison.
Both variants share the same training configuration except for the augmentation strategy used to introduce intra-modal uncertainty.
Replacing spectrogram masking with audio mixing slightly improves A$\rightarrow$T but degrades T$\rightarrow$A performance.
This may be due to a mismatch in the induced uncertainty.
Audio mixing introduces multiple acoustic events, whereas token masking removes part of the caption information, leading to inconsistent semantic representation.
In contrast, combining audio mixing with caption concatenation substantially improves performance across all metrics.
These results suggest that, for audio-language representation learning, modeling uncertainty through information composition via mixing may be more suitable than information removal through masking.

\section{Conclusion}

We propose \shortname, a probabilistic audio-language pretraining framework that replaces masking-based uncertainty with additive uncertainty through audio mixing.
Unlike masking, which destroys semantic content for transient sounds or fails to create hierarchy for ambient sounds, mixing constructs a semantic superset that naturally satisfies the inclusion relationship.
We further introduce a multi-level inclusion loss conditioned on the mixing ratio to promote graded semantic understanding.
Experiments demonstrate improved zero-shot retrieval performance, validating that additive uncertainty better captures the structural properties of audio signals.

\pagebreak

\section{Generative AI Use Disclosure}

The use of generative AI in this paper was limited strictly to text refinement and clarity improvements. The authors are solely responsible for all scientific content, including the methodology, experiments, analysis, and conclusions.

\bibliographystyle{IEEEtran}
\bibliography{longstrings,mybib}

\end{document}